\documentclass[5p]{elsarticle} 

\usepackage{hyperref} %\usepackage{lineno,hyperref}
\usepackage{amssymb}
\usepackage{amsmath}
\usepackage{array}
\begin{document}

\begin{frontmatter}

\title{Energy loss and inelastic diffraction of fast atoms at grazing incidence.}

%% Group authors per affiliation:
%\author{Elsevier\fnref{myfootnote}}
\author{Philippe Roncin}
\author{Maxime Debiossac}
\address{Institut des Sciences Mol\'{e}culaires d'Orsay (ISMO), CNRS, Univ. Paris-Sud, Universit\'{e} Paris-Saclay, Orsay, France}
\author{Hanene Oueslati}\author{Fay\c{c}al Raouafi}
\address{Institut pr\'{e}paratoire aux \'{e}tudes scientifiques et techniques (IPEST), La Marsa, Universit\'{e} de Carthage, Tunisie}

%\fntext[myfootnote]{Since 1880.}

\begin{abstract}
The diffraction of fast atoms at grazing incidence on crystal surfaces (GIFAD) was first interpreted only in terms of elastic diffraction from a perfectly periodic rigid surface with atoms fixed at equilibrium positions.
Recently, a new approach has been proposed, referred here as the quantum binary collision model (QBCM). The QBCM takes into account both the elastic and inelastic momentum transfers via the Lamb-Dicke probability.
It suggests that the shape of the inelastic diffraction profiles are log-normal distributions with a variance proportional to the nuclear energy loss deposited on the surface.
For keV Neon atoms impinging on a LiF(001) surface under an incidence angle $\theta$, the predictions of the QBCM in its analytic version are compared with numerical trajectory simulations. 
Some of the assumptions such as the planar continuous form, the possibility to neglect the role of lithium atoms and the influence of temperature are  investigated. 
A specific energy loss dependence $\Delta E\propto\theta^7$ is identified in the quasi-elastic regime merging progressively to the classical onset  $\Delta E\propto\theta^3$. 
The ratio of these two predictions highlights the role of quantum effects in the energy loss.
%This regime where only one or two binary collision are inelastic is probably easier to identify in the angular scattering width.

\end{abstract}


\end{frontmatter}

\section{Introduction}
The energy loss of keV ions at solid surfaces has been investigated in detail both from the theoretical and experimental points of view. One of the important regimes at low energy is the nuclear regime where electronic excitations play a minor role.
At grazing angle, collisions of keV atoms can be gentle enough to allow scattering in a quantum regime as illustrated by clear diffraction features (see e.g. \cite{Winter_PSS_2011} for a review). 
The identification of elastic fast atom diffraction, characterized by the absence of energy exchange with the surface, was predicted almost ten years ago \cite{Manson2008} but experimental evidences of a well defined Laue circle with diffraction spots size limited by that of the primary beam were scarce and hardly quantified~\cite{DebiossacPRL,Debiossac_Nim_2016,Busch_2012}.

A first attempt to quantitatively describe both the elastic and inelastic diffraction of fast atoms, together with the associated line profiles, was recently proposed \cite{Roncin_PRB_2017}.
The succession of binary collisions with surface atoms along the classical trajectory was described using an idealized trajectory giving close analytic form of the energy loss and inelastic profiles.
Starting from the quantum properties of the individual surface atoms considered as harmonic oscillators during a distant binary collision, the model offers a smooth transition between the quantum and classical regimes with specific predictions on the onset for the nuclear energy loss.

After a short presentation of the QBCM, some of its assumptions are analyzed with more realistic trajectory simulation using the Ne-LiF system to better understand the limitations. Accordingly less importance will be given to an \textit{a priori} justification since these can be discussed in view of the simulations.

    \begin{figure}	\centering \includegraphics[width=0.9\linewidth]{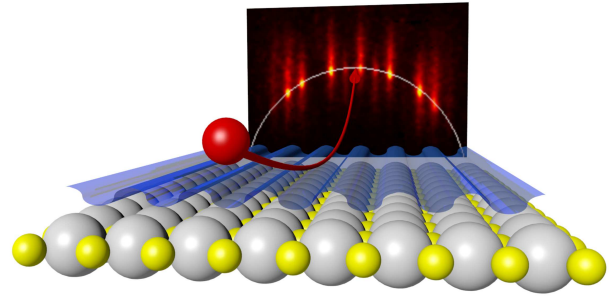}
	\caption{Schematic view of keV Ne atoms impinging on the LiF(001) surface. The diffraction pattern is recorded on an detector \cite{Lupone_2017} placed $\sim 1$m downstream. The elastic diffraction spots are localized on the Laue circle (white line) while the inelastic intensity shows elongated streaks around the Laue circle.
		\label{fig_Setup}} \end{figure}
	
\section{Established theory}	
\subsection{Elastic diffraction; \newline the rigid lattice and the potential energy landscape.}
Theoretical approaches to GIFAD consider the surface as an ideal system with atoms standing still at their equilibrium positions so that the potential energy landscape (PEL) of the helium-surface is perfectly periodic.
The dynamics of the projectile atomic wave-function on this PEL, \textit{i.e.} the diffraction, has been modeled via wave-packet \cite{Rousseau_2007,Aigner_2008}, transition matrix \cite{Manson2008}, semi-classical trajectories \cite{Winter_PSS_2011,Gravielle_2014}, Bohmian trajectories \cite{Sanz}, close coupling \cite{DebiossacPRB_2014,Zugarramurdi_2012} or multi-channels Hartree methods \cite{Diaz_2016b}. 

\subsection{Reduced dimension of the PEL}
Elastic diffraction of fast atoms of energy $E$ impinging on a surface with an incidence angle $\theta_{in}$, was early understood\cite{Rousseau_2007} and described by a 2D problem where an effective particle with perpendicular energy $E_\perp = E \sin\theta_{in}$ evolves in a 2D PEL $V_{2D}(y,z)=\int V_{3D}(x,y,z) dx/d_x$. 
Here $x$ is taken along the low index crystal axis as depicted in Fig.\ref*{fig_Setup} and $d_x$ is the period along x.
This axial channeling approximation (ASCA) was established with quantitative criterion  well satisfied for grazing incidence keV projectile and small lattice units~\cite{Zugarramurdi_2012,Diaz_2016b}. 
Experimentally this is evidenced by the presence of only one Laue circle.

Here we focus on inelastic processes which are described as individual binary collisions along the projectile's travel along $x$, \textit{i.e.} precisely the direction neglected in ASCA.
We assume that the mean properties of these trajectories are well estimated by the trajectory on the mean planar potential defined as 

$V_{1D}(z)=\frac{1}{d_xd_y}\int\int V_{3D}(x,y,z) dxdy$ where $d_y$ is the period along the direction $y$.
Assuming an exponential form for $V_{1D}(z)\propto e^{-\Gamma z}$, the trajectory $z(t)$ is analytic and so are its first and second derivatives $p_z(t)$ and $\dot{p_z}(t)$ describing the momentum transfer to the surface per unit time or unit length. 
Considering that only one atom with mass $m$ per lattice unit $a$ receives the exchanged momentum, an energy deposition curve can be defined and integrated to produce an energy loss~\cite{Manson2008,Rousseau_2008}.
 
\begin{equation}
E_{loss} = \frac{2}{3} \mu E \Gamma a \theta_{in}^3.  \label{Eloss}        
\end{equation}
where $\mu=M/m$ with $M$ the projectile mass and $m$ the mass of a surface atom.
The energy deposition curve has a quasi-gaussian profile \cite{Roncin_PRB_2017} and its full width at half maximum can be used to define the trajectory length $L\propto1/\theta$~\cite{Rousseau_2007}.
This important parameter $L$ can be expressed as the number of most active binary collisions. If one assumes that all collisions are equivalent, the incident beam is deflected by an angle $2\theta_{in}=\theta_{in}+\theta_{out}$, arising from a series of $N_{eq}$ equivalent deflections by $\delta \theta = 2\theta_{in}/N_{eq}$, each of them producing a recoil energy $E_r= \mu E \delta \theta^2$ with
\begin{equation}
N_{eq} =\frac{6}{\Gamma a \theta_{in}}.   \label{Neq}        
\end{equation}
%The QBCM considers two different kinds of momentum transfer, the ones responsible for the specular reflection and the variations around the former value due to the thermal displacement.
%The first ones are calulated with surface atoms at equilibrium position while the second derive from thermally displaced atoms.
%Such distinction appears arbitrary in a classical simulation but if the target atom is considered as an harmonic oscillator the central value of its stationay wave function is well centered at its equilibrium position.

\subsection{Thermal movement of the surface atoms}
At finite temperature the crystal hosts a population of phonon giving rise to movement with an amplitude $\sigma_z^2=\langle   z^2  \rangle $.
 
 Considering the coupled oscillators and averaging over the thermal distribution, the $z$ distribution is gaussian with \cite{Desjonqueres_94}
\begin{equation} \langle   z^2  \rangle  = \frac{3\hbar}{2m\omega}\coth(\frac{T_D}{2T}) =  \frac{3\hbar^2}{2mk_B T_D}\coth(\frac{T_D}{2T}) \label{z2T}; \end{equation}
where $k_B$ is the Boltzmann constant, $T_D$ is the Debye surface temperature so that $\hbar\omega=k_B T_D$ is the energy of a vibration quantum of the local Debye oscillator. 
Interpreting $\sigma_z^2=\langle   z^2  \rangle $ as the variance of the probability to find a surface atoms away from the surface plane, the surface is far from being as flat as idealized in the rigid lattice model.
These displacements induce deviation from the ideal trajectory and affect the coherence of the diffracted signal.

As often encountered in quantum mechanics, the situation can be approached in two ways, either in the real space as summarized below or in the momentum space described in the next section. 
The real space approach considers the coherence of the waves emitted by an ensemble of diffraction centers distributed around their equilibrium positions.
In thermal energies atom scattering or in Xray diffraction, where the scattering takes place on a single atom, this gives rise to the Debye-Waller factor DWF=$e^{-\langle (k_z z)^2 \rangle}\sim e^{-k_z^2 \langle z^2 \rangle}$ for the specular reflection of a wave-vector $k_z$. In GIFAD, the momentum exchange is spread along the successive tiny collisions with the surface atoms and each one only contributes with a dephasing $d\phi=\delta k_z z$ where $\delta k_z$ can be estimated as $\delta k_z=2k_z/N_{eq}$. All these contributions add up incoherently so that the overall dephasing is reduced.
The specific DWF for GIFAD is much more favorable~\cite{Rousseau_2008,Manson2008} $e^{-k_z^2 \sigma_z^2 /N_{eq}}$, indicating that the scattering takes place on a row of $N_{eq}$ active atoms reducing by $\sqrt{N_{eq}}$ the amplitude of the thermal oscillations.

%In thermal energies atom scattering the interaction with a surface atom is dephased by a quantity $d\phi=2k_z z$ corresponding to the specular reflection of a wave vector $k_z$ on a surface atom displaced by a amount $z$. Taking a gaussian distribution of width $\sigma_z$ for the displacement $z$ around its equilibrium position, the phase distribution is gaussian with a width $\sigma_\phi = 2k_z \sigma_z$ giving an overall coherence fraction $\sigma_\phi = 2k_z \sigma_z$ \textit{i.e.} the Debye-Waller factor  DWF=$e^{-k_z^2 \sigma_z^2}$. 
%In GIFAD, the momentum exchange is spread along the successive tiny collisions with the surface atoms and each one only contributes to a dephasing $d\phi=\delta k_z z$ where $\delta k_z$ can be estimated as $\delta k_z=2k_z/N_{eq}$.
%The specific DWF for GIFAD is much more favorable\cite{Rousseau_2008,Manson2008} $e^{-k_z^2 \sigma_z^2 /N_{eq}}$, indicating that the scattering take place on a row of $N_{eq$ atoms reducing the thermal oscillations by $\sqrt{N_{eq}}$.

In both cases, the elastic signal corresponds to atoms at their equilibrium position, and its intensity is attenuated by thermal displacement. 
The fate of the incoherent signal  is however less clear. Where does it appear? Under what conditions the diffraction features remain visible? In other words, how can we describe diffraction pattern in the inelastic regime ?
These questions are easier to address with the momentum approach at the heart of the binary quantum collision model.

%which are used to evaluate the probability that each collision is elastic or not and,

\section{The quantum binary collision model QBCM}
The momentum approach describes the elastic scattering on the surface as a series of elastic collisions with the quantum oscillators. 
If a collision is elastic, the trajectory will again correspond to the classical trajectory associated with the center of the harmonic oscillator \textit{i.e.} as if the rigid lattice description with motionless atoms at their equilibrium positions were real.  
If $q$ is the momentum transferred to this surface atom, then the probability $p_e$ to leave the wave function unchanged is $p_e= |\langle\psi|  e^{iqz} |\psi\rangle|^2$. 
Using the Bloch theorem~\cite{Cohen_Tannoudji} $\langle  e^{iqz}  \rangle = e^{-\frac{1}{2} q^2 \langle   z^2  \rangle  }$, the elastic probability is again the standard DWF.
For an isolated oscillator with pulsation $\omega$ in its ground state the probability is $p_e=e^{-E_r/\hbar\omega}$ with $E_r=\hbar^2q^2 /2m$ the associated recoil energy.
This is equivalent to the Lamb-Dicke probability of recoilless emission which means that in a trapping potential, the wave function may absorb a momentum $q$ without exchanging the recoil energy $E_r$. 
In this respect, $E_r$ is only a virtual recoil energy. 

Taking into account the actual value of $\langle   z^2  \rangle$ on the surface given in Eq.\ref{z2T}, $p_e$ reads
\begin{equation} p_e= exp(-3\frac{E_r}{k_B T_D} \coth(\frac{T_D}{2T}) ). \label{pe} \end{equation}

The product probability $P_e$ that all binary collisions are elastic factorizes to outline the sum of all the virtual recoil energies $E_{loss}=\Sigma_j E_{r\,j}$ along the trajectory

\begin{equation} P_e= exp(-3\frac{E_{loss}}{k_B T_D} \coth(\frac{T_D}{2T}) ). \label{Pe} \end{equation}

If a collision with the surface atom is inelastic, different properly-weighted  initial and final wave functions have to be evaluated.
We can also consider that the momentum dispersion induced by the inelastic collision can be evaluated from classical mechanics with thermally displaced atoms, as if the inelastic collision would project the wave function to its spatial probability distribution $P(z)=|\langle \varPsi| z |\varPsi\rangle|^2$ which is the gaussian distribution of Eq.\ref{z2T}. Since $P(z)$ is centered around the equilibrium value, the median value of the angular distribution $P(\theta_i)$ is the elastic value $\theta_e$ with a, yet unknown, width $\sigma_{\theta_i}$.
Considering these inelastic angular straggling as independent, their contributions are added quadratically for each inelastic collision along the trajectory.
The individual inelastic contributions $\sigma_{\theta_i}$ can be calculated numerically or, within few assumptions on the form of the interaction potential, they can be evaluated analytically. 

\subsection{The QBCM analytic form}

The trajectory and energy loss (Eq.\ref{Eloss}), derived from the exponential planar potential provide a quantitative estimation of the elastic probability associated with the entire trajectory
\begin{equation} DWF=P_e= exp(\frac{- 2\mu E \Gamma a \theta_{in}^3}{k_B T_D} \coth(\frac{T_D}{2T})  ) \label{Pes} \end{equation}
Which is equivalent to the one derived in Ref.\cite{Manson2008} using the mean number of scattering centers $N_{eq}$.

\begin{figure}	\includegraphics[width=0.85\linewidth]{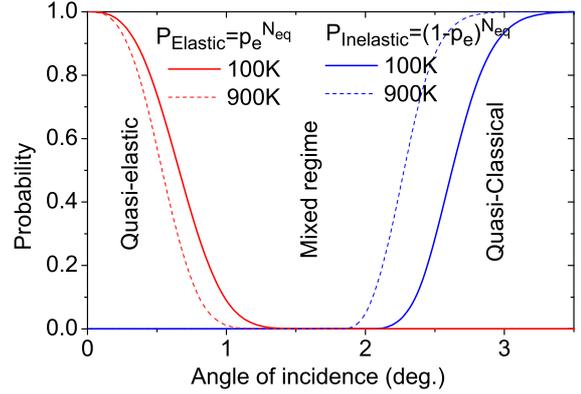}
	\caption{For 1 keV neon atoms, the probability that $N_{eq}= \frac{6}{\Gamma a \theta}$ equivalent collisions are all elastic or inelastic are plotted as a function of the angle of incidence.
		\label{fig_Regimes}}\end{figure} 
The elastic scattering probability is reported in Fig.\ref{fig_Regimes} for different surface temperatures and for a Debye surface temperature of 540 K corresponding to a local harmonic oscillator with $\hbar\omega = k_B T_D$ = 45 meV.

\textit{Stricto sensu}, the equivalent probability $P_{ine}$ that all binary collisions are inelastic is always zero because at large enough distance to the surface the interaction is negligible and $p_{ine}=0$.
Considering only the $N_{eq}$ most important binary collisions, the mean elastic probability $\langle  p_e  \rangle =P_e^{1/N_{eq}}$ is given by

\begin{equation} \langle p_e \rangle = exp(-\frac{\mu E\Gamma^2 a^2 \theta_{in}^4}{3 k_B T_D} \coth(\frac{T_D}{2T}) ) \label{pebarre} \end{equation} 

The complementary mean inelastic probability is $\langle  p_{ine}  \rangle = 1-\langle  p_e  \rangle$. 
This numeric simplification allows a simple estimation of the probability that all $N_{eq}$ binary collisions proceed in the inelastic regime.
These are plotted in Fig.\ref{fig_Regimes} illustrating the progressive merging into the classical regime defined by a unit probability of $P_{inelastic}$.

The exponential form also allows an evaluation of the angular straggling $\sigma_{\theta_i}$ induced by an inelastic transition.
%For an harmonic oscillator, the elastic probability $p_e$ corresponds to the trace of the density matrix of the harmonic oscillator at temperature T.
Assuming an exponential deflection function $P(\theta)\propto e^{-\Gamma z}$, the gaussian position fluctuations $\sigma_z$ of the surface atom are transformed~\cite{Roncin_PRB_2017} into a Log-normal angular distribution of inelastic scattering $P(\theta_i)$. 
As detailed in Table \ref{TabLogNormale},  the elastic scattering value $\theta_e$, corresponding to an in-plane surface atom ($z=0$) is the median value of $P(\theta_i)$.

%\begin{equation} P(\theta_i)=\frac{1}{\sqrt{2\pi}\Gamma\sigma_z\theta_i}exp(\frac{-(\ln \frac{\theta_i}{\theta_e})^2}{2\Gamma^2\sigma_z^2} )  \label{LN} \end{equation}
The variance $\sigma_{\theta_i}^2$ of this inelastic profile is proportional to $\theta_e^2$ (Table \ref{TabLogNormale}) and therefore to the recoil energy  $E_r= \mu E\theta^2$ associated with this tiny deflection;
\begin{equation} \sigma^2_{\theta_i} =\alpha\theta_e^2~=~\alpha\frac{m~E_{r}}{M~E}  \label{sigma_LN} \end{equation} 
with $\alpha=e^{w^2}(e^{w^2}-1)$ and $w=\Gamma\sigma_z$. 

At the end of the trajectory, the angular broadening accumulated in all inelastic events is obtained by adding the variance of the individual contributions $\sigma^2_{\theta}=\Sigma_j \sigma^2_{\theta_j}$ where the sum runs only over the inelastic events.

Expressed in energy $\sigma^2_{\theta}=\frac{m~\alpha}{M~E}~\Sigma_j E_{r\,j}$ 
where the sum is the total inelastic energy loss $\Sigma_j E_{r\,j}$, the angular broadening can be estimated with the actual mean energy loss 
$\langle \Delta E \rangle=\Sigma_j(1-p_{e\,j})E_{r\,j}$ 
where the sum now runs over all scattering events.

Here one has to take into account that the log-normal inelastic scattering profile is asymmetric (Table \ref{TabLogNormale}) because  the excess momentum transfer when an atom is protruding from the surface is larger than the corresponding lack when it recedes.
As a result the elastic scattering angle $\theta_e$ corresponding to the surface atom at its equilibrium position is only the median value.
The peak (the mode) of the inelastic distribution $\theta_{i\,max} = \theta_{e}~e^{-w^2}$ is slightly under-specular while the mean value is over-specular $\langle  \theta_{i}  \rangle = \theta_{e}~e^{w^2 /2}$.

\renewcommand{\arraystretch}{1.7}
\begin{table}
\begin{center}
	\begin{tabular}{| c | c ||c | c | }
		\hline 
	%	\multicolumn{4}{|c|}{Inelastic scattering angle and energy loss profiles} \\
	%	\multicolumn{4}{|c|}{$E_{r}  = \mu E \theta_e^2,~A=\frac{1}{w\sqrt{2\pi}},$ and $w =\Gamma \sigma_z$} \\	\hline	
	    \multicolumn{2}{|c||}{$P(\theta_i)=\frac{A}{\theta_i}e(\frac{-\ln^2 \frac{\theta_i}{\theta_e}}{2w^2} )$} & \multicolumn{2}{c|} {$P(E_{ri})=\frac{A}{2E_{ri}}e(\frac{-\ln^2 \frac{E_{ri}}{E_{r}}}{8w^2} )$}\\ \hline 
	%	$P(\theta_i)$ & $\frac{A}{\theta_i}e(\frac{-\ln^2 \frac{\theta_i}{\theta_e}}{2w^2} )$ &$P(E_{ri})$ & $\frac{A}{2E_{ri}}e(\frac{-\ln^2 \frac{E_{ri}}{E_{r}}}{8w^2} )$ \\ \hline 
		$\langle \theta_i \rangle$ & $\theta_e e^{w^2 /2}$ &$\langle E_{ri} \rangle$ & $E_r e^{2w^2}$ \\ \hline
		mode & $\theta_e e^{-w^2}$ &mode & $E_r e^{-4w^2}$ \\ \hline
		$\sigma_{\theta i}^2$ & $\theta_e^2 (e^{w^2}-1)e^{w^2}$ &$\sigma_E^2$ & $E_{r}^2(e^{4w^2}-1)e^{4w^2}$ \\ \hline
		%	\multicolumn{4}{|c|}{ $A=\frac{1}{w\sqrt{2\pi}},$ $E_{r}  = \mu E %\theta_e^2$ and $w =\Gamma \sigma_z$} \\\hline
	\end{tabular}
\end{center}
\caption{Properties of the log-normal polar scattering angular and energy loss distributions associated with an inelastic event.\newline	
	 $E_{r}  = \mu E \theta_e^2,~A=\frac{1}{w\sqrt{2\pi}},$ $w =\Gamma \sigma_z$}.
\label{TabLogNormale}
\end{table}
%Table 1. properties of the log-normal distributions associated with an inelastic event. $E_{r}  = \mu E \theta_e^2,~A=\frac{1}{w\sqrt{2\pi}},$ $w =\Gamma \sigma_z$.
Accordingly, the individual inelastic energy loss distribution is also asymmetric with a log-normal width $w'=2w$ due to the square dependence $E_{ri} =\mu E\,\theta_i^2$.
The integrated energy loss where all collisions are inelastic given in in Eq.\ref{Eloss} rewrites ;
\begin{equation}
E_{loss} = \frac{2}{3} \mu E \Gamma a \theta_{in}^3~e^{\Gamma^2\sigma_z^2} \label{Eloss_w2}        
\end{equation}
Compared with the standard rigid lattice energy loss in Eq.\ref{Eloss}, the classical limit of the inelastic energy loss above is now temperature dependent because the mean thermal amplitude $\sigma_z^2=\langle z^2 \rangle$ in Eq.\ref{z2T} is part of the log-normal asymmetric profile via $w=\Gamma\sigma_z$.

The actual energy loss $\Delta E=\Sigma_j(1-p_{e\,j})E_{r\,j}$ can be integrated numerically or estimated with $N_{eq}$ (Eq.\ref{Neq}). Using $\langle p_e \rangle =P_e^{1/N_{eq}}$ and $\langle E_r \rangle = E_{loss}/N_{eq}$,  
$\Delta E\sim N_{eq} (1-\langle p_e \rangle) \langle E_r \rangle = (1-\langle p_e \rangle) E_{loss}$, or explicitly,

\begin{equation}
\Delta E \sim (1-\langle p_e \rangle)~ \frac{2}{3} \mu E \Gamma a \theta_{in}^3~e^{\Gamma^2\sigma_z^2}\label{Eloss_pond}  
%\Delta E \sim [1-exp(-\frac{\mu E\Gamma a \theta^3}{k_B T_D} \coth(\frac{T_D}{2T}) ) ]~ \frac{\mu E\Gamma^2 a^2 \theta^4}{9} ~e^{\Gamma^2\sigma_z^2} \label{Eloss_pond}        
\end{equation}
with $\langle p_e \rangle$ taken from Eq.\ref{pebarre}.
We will see below that $\Delta E$ can be much smaller than the sum of the recoil energies in Eq.\ref*{Eloss_w2} obtained without taking into account the Lamb-Dicke factor.

The surface temperature enters both the quantum probability $\langle p_e \rangle$ with the term $\coth(\frac{T_D}{2T})$ and the classical mean inelastic energy loss with the $e^{\Gamma^2\sigma_z^2}$ factor. Note that the inelastic scattering angle and energy loss are correlated so that, for a given angle of incidence $\theta_{in}$, the energy loss depends on the polar scattering angle \cite{Roncin_PRB_2017} as observed in Ref.\cite{Villette_these} and  Ref.\cite{Winter_2002}.

The QBCM indicates a direct connection between the actual energy loss and the angular straggling.
For quantitative comparison, the line broadening taking place along the $y$ and $z$ direction is averaged over all impact parameters as detailed in Ref.\cite{Roncin_PRB_2017}.

%\begin{equation}
%\fbox {$
%	\begin{aligned}
%inelastic~polar~angle~\theta_i~&,~~energy~loss~E_{ri}  = \mu E \theta_i^2\\
%	P(\theta_i) =\frac{A}{\theta_i}e(\frac{-\ln^2 \frac{\theta_i}{\theta_e}}{2w^2} )~~&,~~P(E_{ri}) =\frac{A}{2E_{ri}}e(\frac{-\ln^2 \frac{E_{ri}}{E_{r}}}{8w^2} )\\
%	\langle \theta_i \rangle = \theta_e e^{w^2 /2}~~&,~~\langle E_{ri} \rangle = E_r e^{2w^2}\\ 
%	mode  = \theta_e e^{-w^2}~~&,~~mode = E_r e^{-4w^2}\\
%	\sigma_{\theta_i}  = \theta_e \sqrt{(e^{w^2}-1)e^{w^2}}~~&,~~ \sigma_{E} = E_{r} \sqrt{(e^{4w^2}-1)e^{4w^2}} \\
%	\end{aligned}
%	$}     
%\label{LogNormale}
%\end{equation}
%$E_{r} \sqrt{(e^{4w^2}-1)e^{4w^2}}$
\subsection{The quasi-elastic form factor}

Though beyond the scope of this paper, it seems difficult to bypass the  following question : Is there a form factor in inelastic diffraction? In other terms is it just a stochastic degradation of the Laue circle intensity or can we use the inelastic intensity to retrieve information of the shape of the PEL?

Experimentally, the sharp increase of the scattered intensity at the Laue circle was not correlated to a marked variation of the relative intensities.
A smooth continuity is always observed on either side of the Laue circle indicating a smooth continuity of the associated form factor.
This was investigated quantitatively~\cite{Roncin_PRB_2017} using the hard corrugated wall model (HCW) to fit all the relative intensities $I_m = J_m (2k_\perp z_c)^2$ with $J_m$ the Bessel function of rank $m$ and $z_c$ the sinusoidal corrugation amplitude as the only free fitting parameter.
The values of $z_c$ derived from experimental elastic or inelastic intensities are almost identical provided that one uses the effective wave vector $k_{\textsf{eff}} = (k_{in} + k_{out})/2$ in the HCW formula above. Note that the good agreement is observed only on a limited angular range around the Laue circle, typically within one standard deviation $\sigma_{\theta_i}$.

In the quasi-elastic regime where mainly one inelastic collision takes place, the trajectory can be splitted into two parts, the way in and the way out.
Since the inelastic event is most likely located close to the turning point where the momentum transfer is maximum, the two separate diffraction events correspond to two half diffraction event, one with a momentum $k_{in}$ and one with a momentum $k_{out}$. 
This can be qualitatively understood using again the HCW model where the intensity modulation comes from a path difference between trajectories on top of a row or in between resulting in a phase difference $2k_{in} z_c$. % where $z_c$ is the corrugation amplitude. 
Here, the phase difference naturally splits into two terms corresponding to the incoming and outgoing wave-vectors adding up to $(k_{in} + k_{out}) z_c$ defining  the effective wave-vector $k_{\textsf{eff}} = (k_{in} + k_{out})/2$.

\subsection{3D trajectory simulations} 

The closed form given in Eq.\ref{Pes} and Fig.\ref{fig_Eloss_1} were obtained from an analytic form of the atom trajectory. 
This simplified trajectory is consistent with the ASCA where the contribution of individual atoms is replaced by the average over the lattice unit along $x$. 
However, the analytic form of the trajectory is obtained with specific limitations on the form of the binary interaction potential and with only one atom per lattice site.
The 3D trajectory simulations has no restriction on the form of the binary potential nor on the structure or composition of the lattice unit.
We describe here such 3D trajectory simulations to investigate some of the approximations.
We use conditions where the continuous model is supposed to be valid.

%The analytic version presented above tries to capture the mean behavior of a real surface by the behavior on the mean planar interaction potential.

One of the underlying assumptions of the analytic form is that the deflection accumulated along a distance $a$ corresponding to the lattice unit can be attributed to only one atom of the unit cell.
This aspect can be tested by constructing the interaction potential as a sum of binary interaction potentials allowing each surface atom to receive the exact amount of momentum transfer.
Using binary interaction potentials is not, \textit{a priori}, a very severe restriction; it can be seen as a convenient and compact description of the potential energy landscape (PEL) calculated for instance from density functional theory and usually on a rigid lattice.
These effective binary interaction potentials fitted on the PEL can then be used to estimate the new PEL associated with displaced atoms forming a non periodic arrangement, more difficult to calculate \textit{ab initio}.

The simulation is based on a classical trajectory program developed to reproduce scattering profiles recorded with fast ions or atoms~\cite{Villette_2000,Pfandzelter}.
The surface atoms do not have time to move during the collision and are taken immobile.
For standard classical scattering calculations, the atoms are usually thermally displaced~\cite{Villette_2000,Pfandzelter}. Here, according to the QBCM, the atoms are described by their harmonic oscillator wave-functions whose central position is the equilibrium position of the rigid lattice.
For a given angle of incidence and impact parameter the 3D newton equation is integrated with a 4$^{th}$ order Runge-Kutta method.
The original program focused only on the projectile trajectory~\cite{Villette_these}, but it has been adapted here to track the momentum $\delta k_{ij}$ transfered to each surface atom labeled by the miller indexes $i,j$.
At the end of the trajectory, the individual elastic probabilities $p_{e\,ij}$ can be estimated from Eq.\ref{pe} for arbitrary Debye surface temperature and crystal temperature.
For each trajectory, the classical energy loss ($E_{loss}$ above) is simply defined as the sum of the recoil energies $E_{r\,ij}=\delta k_{ij}^2/2m$ while the actual mean energy loss ($ \langle   \Delta E  \rangle$ above) is weighted by the inelastic probability $(1-p_{e\,ij})E_{r\,ij}$.
When summed over all different trajectories along a given direction of the surface, the mean energy loss and the mean elastic fraction are determined.

\subsection{Binary interaction potentials}

The form used here to expand the binary interaction potential is the screened coulomb form $V(r)=\Sigma_j~V_j(r)$ with $V_j(r)=a_j/r~e^{-r/b_j}$ used by Yukawa, Moli\`{e}re, Ziegler-Biersack-Littmark \cite{ZBL_85}, O'Connor-Biersack~\cite{OCB_86} and many authors~\cite{Winter_2002}.
Using a sufficient number of terms this expansion is able to represent realistic potentials with different contributions such as shell effects, polarization or Van der Waals attractive terms \cite{Bocan_2016}.
Here the purpose is to examine the assumption of the QBCM in its planar continuous form which assumes that the atom-surface interaction potential is well represented by an exponential form $V(z)=V/A~e^{-z/r_c}$ where $r_c=1/\Gamma$ is the effective range and $A$ the area of the unit cell.
Using a single atom per lattice unit and a single Yukawa term having the same $r_c$ value produces a mean planar potential $V_{1D}(z)$ with the correct asymptotic exponential form. This was the motivation for using this form.

% and where the prefactor $V$ is not important here as it only shift the whole trajectory in $z$ without affecting the rate of momentum transfer.
%The actual distance of the turning point will only be important when evaluating the line-shapes of inelastic diffraction\cite{Roncin_PRB_2017}.

If one were to choose the best $r_c$ value for a given problem, one should probably try to adjust the prefactor and the range, such that the single Yukawa term be more or less tangent to the real potential close to the turning point where most of the momentum transfer takes place. Here we simply consider the single Yukawa potential used in Ref.\cite{Villette_2000} and Ref.\cite{Khemliche_2001} to study the grazing scattering of Ne$^+$ ions on LiF;
$V_F(r)=4400 eV/r~e^{-1.9r}$ and $V_{Li}(r)=13.4 eV/r~e^{-2.2r}$ with $r$ in $a_0$ (0.53 $\AA{}$).
%Here, to allow straightforward comparison between the model and the simulation we chose to consider only one term in the expansion of the Fluorine and Lithium ions with neon $V_F(r)=a_i/r~e^{-r/b_i}$ and $V_i(r)=a_i/r~e^{-r/b_i}$ \cite{Villette_2000}. 

These coefficients are relatively close to the leading repulsion term considered in Refs\cite{Winter_2002,schueller_2008,Gravielle_2011,Miraglia_2017}.
Before comparing with the continuous model it should be noted that the ASCA is also well confirmed in such a classical trajectory description. 
Whatever the impact parameter within the unit cell, all trajectories end up on the Laue circle within 10$^{-9}$ deg. 
More precisely the scattering angles are insensitive to the initial coordinate $x$ taken here along the crystal axis where the beam is aligned. In fact the ASCA was first derived in a classical trajectory context~\cite{danailov}.
%The same deflection functions $\theta (y)$ and $\phi (y)$ which qualitatively explain the diffracted intensities, could be calulated with axial potential. 
%However, to evaluate in detail the momentum transfer and inelastic contributions, the full 3D calculation is performed.

\section{Results of the 3D trajectory simulations} 
The quantum collision model developed above with the continuous planar potential and analytic trajectory gives compact forms for the elastic scattering probability, energy loss and angular straggling. 
In this section, it is compared with full 3D simulations on the rigid lattice to evaluate its accuracy and variability along different crystallographic axis and to evaluate the assumption of the planar model that Li atoms play a minor role.
For each direction and incidence angle, the elastic probabilities evaluated for each impact parameter as a product of all elastic probabilities $P_e=\Sigma_{ij} ~p_{e\,ij}$ along the trajectory are averaged over the impact parameter. 
%The curves reported correspond to values average over all impact parameters within a unit lattice.
%The surface atoms are immobile but the elastic probability associated with a given recoil energy is evaluated from the temperature dependent formula \ref{pe}.

\subsection{Along the $[100]$ direction}

\begin{figure}	\includegraphics[width=0.85\linewidth]{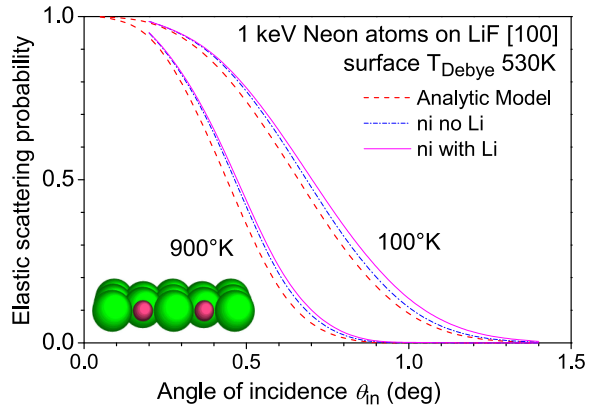}
	\caption{The elastic scattering probability from the continuous 1D model is compared with the 3D numerical integration (ni) with and without taking into account the Lithium atoms.
		\label{fig_cmp_Pe}}\end{figure} 
	
%First, to compare parameters one by one, only Fluorine atoms are considered as in the analytic form of the continuous planar potential.
The elastic scattering probabilities integrated along the $[100]$ direction for which lithium atoms are partly hidden behind the fluorine, are reported in Fig.\ref{fig_cmp_Pe}.

%The one reported as analytic model in (long-dashed red) corresponds to the continuous planar model taking only the leading repulsion term associated with fluorine atoms.
%The rigid lattice calculations correspond to full 3D trajectories with surface atoms at equilibrium positions neglecting or taking into account the lithium ions. 
%The elastic probability reported is evaluated as a product of individual elastic probabilities. 
%These are calculated from the classical recoil energy transferred to each surface atom along each complete projectile trajectory. 

%The dotted blue curves are for the rigid lattice numerical integration with or without lithium ions. 
The 3D simulations with or without taking into account the lithium ions are compared, in the quasi-elastic regime, with the 1D planar model at surface temperatures of 100K and 900K.
All three approaches agree reasonably well with the lowest elastic probability given by the analytic formula.
In this formula the momentum transfer per lattice unit is entirely attributed to a single fluorine atom which has therefore a larger probability to undergo an inelastic transition.
Although weak, the small repulsion from neighbor atoms weaken the recoil and increase the elastic probability.
The same holds for the presence of lithium atoms. At low angle of incidence their contribution is not visible, but close to one degree the lithium atoms contribute to the repulsion, weakening the fluorine recoil energy.
At even larger energy they could even start to be responsible for inelastic transitions but this is not observed in the present range.

The nice agreement between the analytic and 3D simulations with and without lithium is encouraging but the $[100]$ direction is quite favorable with the lithium ions well protected behind the fluorine ions.

\subsection{Along the $[110]$ direction}

\begin{figure}	\includegraphics[width=0.85\linewidth]{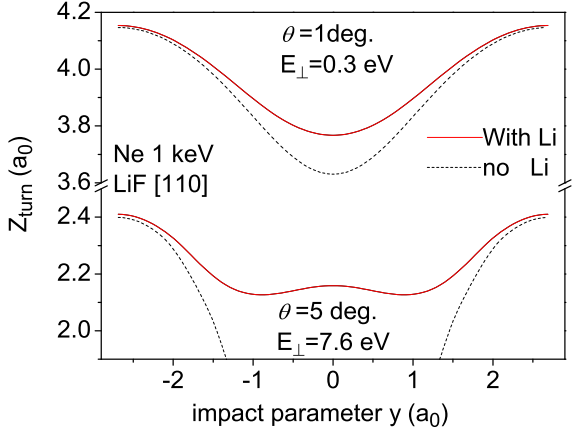}
	\caption{The corrugation function $Z_c (y)$ along the $[110]$ is displayed for $E_\perp$=0.3 and 7.6 eV. The row of Li atoms in the center become distinctly visible only at comparatively large values of $E_\perp$.
		\label{fig_zturn_110}}\end{figure}
\begin{figure}	\includegraphics[width=0.85\linewidth]{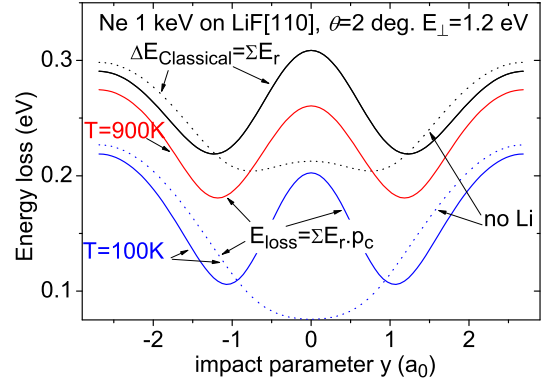}
	\caption{Energy loss as a function of the impact parameter along the $[110]$ direction. The solid curves take the lithium ions into account. The blue and red one converge to the classical prediction in black (reported here for 100K). Dotted curves are without Lithium ions.
		\label{fig_Eloss_110}}\end{figure}

Along the $[110]$ direction, the lithium ions and the fluorine ions form separate rows leaving the Li ions exposed.
The lithium ions have a direct influence on the 2D PEL as illustrated in Fig.\ref{fig_zturn_110} where the corrugation function $Z_c(y)$, represented by the turning point of the trajectories~\cite{Winter_PSS_2011}, is plotted as a function of the impact parameter ($y$). 
Of course, neglecting the lithium ions has drastic consequences on the surface corrugation, already at 300 meV the corrugation increases by 0.2 $a_0$ (0.1 $\AA$) and this would completely modify the observed diffracted intensities.
As a reference, a tiny rumpling of 0.05 $\AA$ of the lithium ions with respect to fluorine could be estimated from the helium elastic diffraction data~\cite{Aigner_2008}.

The question addressed here is different, do the lithium ions play a role in decoherence and whether or not momentum transfer and energy loss are sensitive to the orientation of the surface.
Fig.\ref{fig_Eloss_110} reports the energy loss $\Delta E=\Sigma_{ij}(1-pe_{ij})E_{r\,ij}$ integrated along the trajectory as a function of the impact parameter $y$.
It shows that, at two deg. incidence corresponding to an energy $E_\perp$ of 1.2 eV, the magnitude of the energy loss associated with trajectories on top of a lithium row or fluorine row are comparable. At this angle, the lithium ions contribute to half of the energy loss and this contribution was found to increases with $E_\perp$.

%The Fig.\ref{Eloss} also shows that this contribution depends on the temperature with a lithium contribution slightly less than one half at a surface temperature of 100K and slightly more at the classical limit corresponding to $\Delta E=E_{loss}=\Sigma_{ij}Er_{ij}$ \textit{i.e.} to unit inelastic probability for each binary collision or to infinite temperature. 
%Note that this small effect can not be due to lithium atom depassant more above the surface than fluorine because the model assumes a rigid lattice. 
%Here it could be due to the fact that the lithium ions mainly contribute close to the turning points and that their range could increases faster with temperature than the one of the fluorine atoms.    

\begin{figure}	\includegraphics[width=0.85\linewidth]{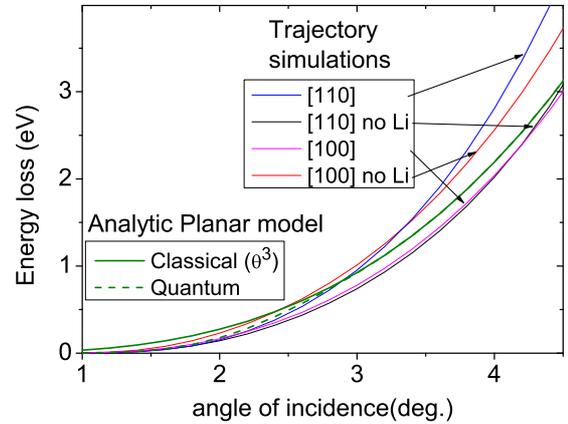}
	\caption{Angular dependence of the energy loss along the $[100]$ and $[110]$ direction and with or without including the Li atoms. The planar model does not include Li atoms.
		\label{fig_Eloss_1}}\end{figure}
 
The mean energy loss, averaged over the impact parameter, is reported in Fig.\ref{fig_Eloss_1} for the $[110]$ and $[100]$ directions and with or without lithium ions. 
At incidence angle above 2 deg. opposite dependencies are observed. Along the $[100]$ directions where lithium ions are hidden, they continue to have a protective role by taking away some momentum elastically. The effect is opposite along the $[110]$ direction where they are directly exposed where the momentum that they exchange with the projectile is large enough to quit the Lamb-Dicke regime.
Improving the planar model beyond the quasi elastic regime is beyond the scope of this paper but the Fig.\ref{fig_Eloss_1} indicates that that Li atoms have a minor contribution below two deg. incidence. 

\begin{figure}	\includegraphics[width=0.80\linewidth]{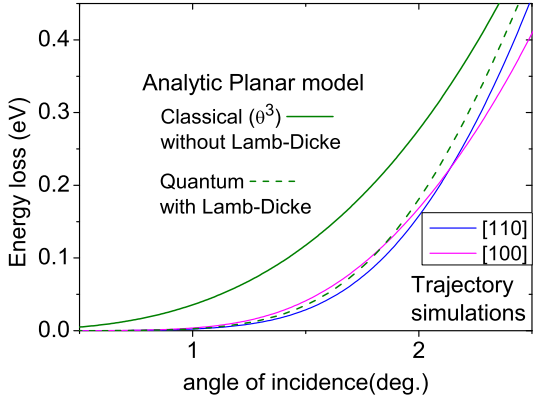}
	\caption{Same as Fig.\ref{fig_Eloss_1} plotted with an enlarged scale to illustrate the effect of the Lamb-Dicke effect neglected in the classical planar model. 
		The Lamb-Dicke effect allows momentum transfer without energy exchange reducing drastically the energy loss at low incidence.
		\label{fig_zoom_Eloss}}\end{figure}

A zoom of the low energy region is displayed in Fig.\ref{fig_zoom_Eloss} showing clearly that the classical formula neglecting the Lamb-Dicke effect, \textit{i.e.} considering $p_e=0$ so that the virtual recoil energy becomes real, is way above the models that take it into account.
	
We plot the same data in log-scale in Fig.\ref{fig_lnEloss}, to clearly show that this quantum effect gives rise to an energy loss scaling with $\theta_{in}^7$ whereas the classical scaling of $\theta_{in}^3$ is obtained when it is neglected.

\subsection{The quasi-elastic energy loss}

\begin{figure}	\includegraphics[width=0.85\linewidth]{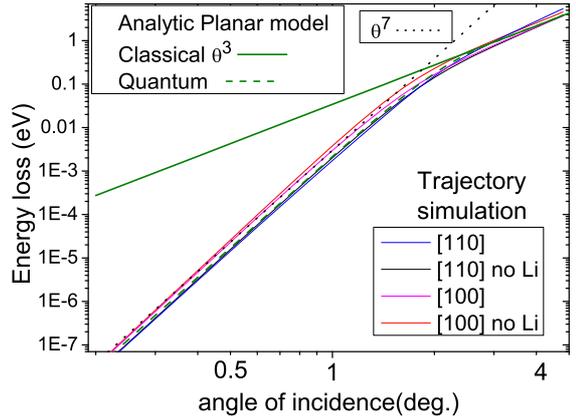}
	\caption{Same as Fig.\ref{fig_Eloss_1} plotted in log-scale to highlight the $\theta^7$ dependence of the quantum energy loss in the quasi-elastic regime.
		\label{fig_lnEloss}}\end{figure}

The $\theta_{in}^7$ scaling observed in Fig.\ref{fig_lnEloss} can be understood in the planar model using the equivalent scatterers simplification where the virtual energy loss $E_{loss}\propto\theta^3$ of Eq.\ref*{Eloss} is assumed to be equiparted among $N_{eq}$ collisions as in Eq.\ref*{Neq} and Fig.\ref{fig_Regimes}. 

This yields an individual recoil energy $E_r\propto\theta^4$ and an individual elastic probability $p_{e} = e^{-\beta\theta^4 }$. 

In the quasi-elastic regime the elastic probability $p_e$ is close to one \textit{i.e.} $p_e\sim 1- \beta \theta^4$ (Eq.\ref{pebarre}) so that the complementary inelastic probability $p_{ine} \sim \beta \theta^4$. 

The mean inelastic energy loss is the weighted sum $\Delta E$:

$\Delta E = N_{eq}.p_{ine}.E_r$ so that $\Delta E~\propto~\theta^7$ 

More precisely combining Eq.\ref{Eloss}., Eq.\ref*{Neq}. and Eq.\ref*{pe}., the asymptotic energy loss $\Delta E_{Quasi-elastic}$ specific of the quasi elastic regime is :
\begin{equation} \Delta E_{Quasi-elastic}= \frac{2\mu^2 E^2 \Gamma^3 a^3}{9~k_B T_D}  \coth(\frac{T_D}{2T}) \theta^7~\label{theta7}. \end{equation}

\begin{figure}	\includegraphics[width=0.85\linewidth]{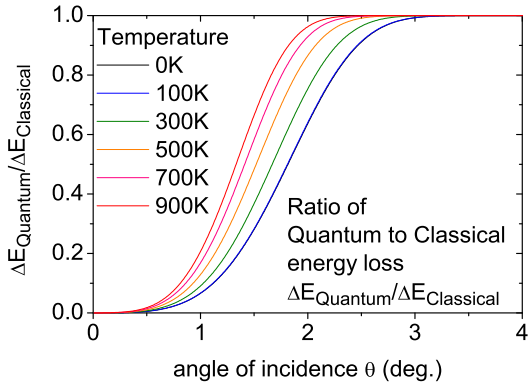}
	\caption{Ratio of the quantum to classical energy loss.
		\label{fig_Eloss_ratio}}\end{figure}

The transition from the quasi-elastic to the classical regime is better illustrated when plotting the ratio of the associated energy loss at different temperatures (Fig.\ref*{fig_Eloss_ratio}). 
It provides a simple illustration of the Lamb-Dicke effect taking into account the quantum nature of the surface. 
Note that the $e^{w^2}$ factor due to the asymmetric energy-loss profile (Eq.\ref{pebarre}) does not enter this ratio.
Compared with the three regimes identified in Ref.\cite{Roncin_PRB_2017} and in Fig.\ref{fig_Regimes}, the ratio allows a simpler separation. It is almost zero in the quasi-elastic regime with an asymptotic form at low incidence, 
\begin{equation} \frac{\Delta E_{Quantum}}{E_{loss}}= \frac{\mu E \Gamma^2 a^2}{3~k_B T_D}  \coth(\frac{T_D}{2T})  \theta^4~\label{ratio} \end{equation}
The ratio has intermediate values in the mixed regime where both elastic and inelastic values become significant before reaching unity in the quasi-classical regime.

The magnitude of the energy loss is rather small, making experiments quite difficult. However the model indicates a direct correspondence of the energy loss with the angular straggling and the elastic fraction.
Provided the surface quality is large enough to reduce the contribution of defects to a negligible value, this energy scaling could be observed in the angular domain.

Another output of the present simulations is that, in the quasi-elastic regime the energy loss is comparable along the $[100]$ and $[110]$ directions as visible on Fig.\ref{fig_zoom_Eloss}. 
This is consistent with the observation by Seifert \textit{et al.}\cite{Seifert_2015}, that the transverse line broadening of helium atoms colliding on mono-layer $SiO_2$ film on $Mo(112)$ and for H atoms colliding on $LiF(001)$ is independent of the crystal orientation. 
In Ref.\cite{Roncin_PRB_2017} this line broadening was assumed to depend mainly on the energy loss $\Delta E$ so that the results on Fig.\ref{fig_zoom_Eloss} link the observation of \cite{Seifert_2015} to the QBCM.
At larger angles of incidence, as illustrated in Fig.\ref{fig_Eloss_1}, the energy loss along both directions start to show significant differences.

\section{Summary and Conclusion}

Compared with the transition matrix approach~\cite{Manson2008} where the whole crystal and all its phonon modes are taken into account, the QBCM considers that each of the successive collisions with a surface atoms is fast enough to prevent any movement at this timescale. 
The binary quantum collision model considers the collisions with the surface atoms as collisions with harmonic oscillators and the main quantum effect associated with this description is the temperature dependent Lamb-Dicke effect describing elastic and inelastic collisions. 

In the quasi-elastic or Lamb-Dicke regime where the individual elastic probabilities are close to one and where the overall elastic probability exceeds a few percent, the continuous planar model compares well with the quantum rigid lattice model with or without lithium atoms. 
In this regime, the inclusion of lithium ions increases slightly the overall elastic scattering probability as these ions take a little share of the projectile momentum and remain in their quantum state. 
This effect is more pronounced along the $[110]$ direction where the lithium ions are not hidden and less protected by the fluorine ones. 
This suggests that the simple planar model, with its handy analytical formulas, is a fair approximation of the quantum rigid lattice model and that it could be used as a model to test the link with energy loss.

%The analytic version presented in Ref\cite{Roncin_PRB_2017} was trying to capture the mean behavior of a real surface by the behavior on the mean planar interaction potential.
%At least in the quasi-elastic regime, the trajectory simulations along the $[100]$ and $[110]$ directions indicate that the simple planar model should provide a fair estimate of the elastic probability and of the energy loss.
 
%\textit{The situation is probably different in the quasi classical regime where most of the individual inelastic probabilities are close to one but where, statistically few collisions are elastic reducing the energy loss and angular straggling.}

This paper draws a link with the ongoing effort to mimic solid state physics with trapped atoms in optical lattices and where an important step is to reach the Lamb-Dicke regime of recoilless emission~\cite{Westbrook_95}. It also provides an opportunity to investigate the onset of the energy loss regime where quantum effects are still important before merging into more classical descriptions of the nuclear contribution and entering again a quantum description of the electronic energy loss see. See Ref.\cite{Bauer_2005,Koval_2017} for recent papers.

The quantum Monte-Carlo approach used in \cite{Aigner_2008} could probably be adapted to describe both the elastic and inelastic scattering by introducing the Lamb-Dicke probability. 
As it was not included, the model was not able to predict the elastic scattering and the inelastic angular profiles and line-widths had to be adjusted \textit{ad hoc}. 
Also, the QBCM suggests that the PEL landscape to be considered is not the one corresponding to the thermal average~\cite{Aigner_2008} but the one of the rigid lattice. 
The accuracy on the rumpling of the lithium atoms should be reevaluated accordingly~\cite{Schuller_PRA2010}.

Compared with the former binary collision approximation (BCA), the QBCM takes into account the quantum nature of the surface atoms. 
The low energy limit of the predicted energy loss is therefore profoundly different as illustrated by the $\theta^7$ dependence. 
More unexpected, the QBCM also predicts a different temperature dependence. 
In Ref.\cite{Winter_2002} both experimental results and numerical simulations with thermally displaced atoms show that the energy loss increases with temperature, a feature not reproduced by the standard rigid lattice model neither by the classical formula in Eq.\ref{Eloss}.
In the QBCM, Eq.\ref{Eloss_w2} predicts a marked increase of the mean energy loss with the surface temperature due to the asymmetric log-normal energy loss profile.
In the present form, the rigid lattice is significantly more predictive than in the classical the BCA, both in the low energy and high temperature limits.
The predictions have to be compared with more extensive trajectory simulations and with experiments to explore the range of validity and understand the limitations.

\section{Perspectives}
%The QBCM two options, one is to average all scattering properties over the lattice unit

Just like neutrons, atoms have a comparatively large mass so that inelastic diffraction always takes a significant part of the intensity.
Understanding its origin and consequences will allow better interpretations of grazing incidence fast atoms diffraction with solid surfaces.
The inelastic description should continue progressing with more complex inorganic surfaces such as vicinal surfaces \cite{Desjonqueres_2002} or oxide layers \cite{Feiten_2015}. 
The case of single layers, organic~\cite{Seifert_alanine} or inorganic~\cite{Debiossac_PRB_2016,Zugarramurdi_15,Seifert_2010} is particularly interesting because the Debye frequency should be related to the coupling to the surface which is a very important parameter.

With thermal helium inelastic diffraction, this binding has been identified with spectroscopic resolution by specific acoustic phonon modes \cite{Taleb_2016}.
Measurements with fast atoms will probably never reach such sensitivity but the information is to be searched in the line-shape of inelastic diffraction spots. Recent investigation of inelastic diffraction of thermal energies neon atoms \cite{Taleb_2017} shows striking similarities with the log-normal profiles discussed here and in Ref.\cite{Roncin_PRB_2017}, suggesting a possible link between the detailed phonon approach and the simplified Lamb-Dicke approach. 

Inelastic diffraction could also help in achieving a better description of the  atomic triangulation method \cite{Feiten_2015,Nataliya}.
In this new technique, only the overall momentum transfer in the polar and azimuthal directions are investigated, in the same range of polar incidence, but as a function of the azimuthal angle of the surface \cite{Sereno_2016}.
This allows for high contrast identification of the directions where the molecules tend to align on the surface and to form rows.
The diffraction, if present, is not analyzed in detail but quantum treatment of the scattering properties around the channeling direction \cite{Zugarramurdi_2013,Debiossac_PRA} could be more efficient than classical treatments and inelastic effects could be the missing link between classical and quantum descriptions.
Last but not least, the use of fast atom diffraction inside a molecular beam epitaxy vessel has proven to be very effective~\cite{Debiossac_ASS_2017}, both to characterize freshly prepared surfaces~\cite{DebiossacPRB_2014} and to follow the growth process \cite{Atkinson_2014}. The account of inelastic scattering should open new applications such as temperature sensing, defect sensitivity.

\end{document}